\documentclass[pra,twocolumn,a4paper,showpacs,superscriptaddress]{revtex4}

 \usepackage{latexsym}
 \usepackage{amsmath}
 \usepackage{amsfonts}
 \usepackage{graphicx}
 \usepackage{amssymb}
 \usepackage{subfigure}
 \usepackage{color}
 \usepackage{mathtools}

 \newcommand{\ket}[1]{\ensuremath{|#1\rangle}}
 \newcommand{\bra}[1]{\ensuremath{\langle #1 |}}

 \newcommand{\bc}{\begin{center}}
 \newcommand{\ec}{\end{center}}

 \newcommand{\ii}{i}
 \newcommand{\DP}{\Delta_p}

 \newcommand{\bl}{\beta_{L}}

 \newcommand{\bs}{\beta_{T}}

 \usepackage[utf8]{inputenc}

\begin{document}

\title{Phase induced transparency mediated structured beam generation in a closed-loop tripod configuration}

\author{Sandeep \surname{Sharma}}
\email{sandeep.sharma@iitg.ernet.in}
\affiliation{Department of Physics, Indian Institute of Technology Guwahati,
Guwahati- 781039, Assam, India}

\author{Tarak N. \surname{Dey}}
\email{tarak.dey@iitg.ernet.in}
\affiliation{Department of Physics, Indian Institute of Technology Guwahati,
Guwahati- 781039, Assam, India}
\date{\today}

\pacs{42.50.Gy,42.65.-k, 42.50.Tx}
\begin{abstract}
We present a phase induced transparency based scheme to generate structured beam patterns in a closed four level atomic system. 
We employ phase structured probe beam and a transverse magnetic field (TMF) to create phase dependent medium susceptibility. 
We show that such phase dependent modulation of absorption holds the key to formation of a structured beam. 
We use a full density matrix formalism to explain the experiments of Radwell et al. [Phys. Rev. Lett. 114, 123603 (2015)] at weak probe limits.
Our numerical results on beam propagation confirms that the phase information present in the absorption profile gets encoded on the spatial probe envelope which creates petal-like structures even in the strong field limit. 
The contrast of the formed structured beam can be enhanced by changing the strength of TMF as well as of the probe intensity. 
In weak field limits an absorption profile is solely responsible for creating a structured beam, whereas in the strong probe regime, both dispersion and absorption profiles facilitate  the generation of high contrast structured beam.
Furthermore we find the rotation of structured beams owing to strong field induced nonlinear magneto-optical rotation (NMOR).
\end{abstract}
\maketitle
\section{Introduction}
In the recent past, there has been a growing interest in the creation of structured beams due to their potential applications in optical micromanipulation \cite{Woerdemann}, quantum information processing \cite{Vaziri}, microtrapping and alignment \cite{Gahegan,Padgett,Macdonald, Paterson}, and the biosciences \cite{Stevenson}. 
Various techniques for the generation of such beams exists that uses conventional optical components like digital micro-mirror device(DMD) \cite{Zupancic}, laser resonator \cite{Schwarz, Naidoo}, axially symmetric polarization element \cite{Sakamoto}, porro-prism \cite{Burger, Litvin} and spatial light modulator \cite{Jesacher, Schmitz, Arnold}.
In most of the techniques the key feature is to coherently superpose two Laguerre-Gaussian(LG) beam modes with equal but opposite orbital angular momentum(OAM), thus creating a structured beam profile.
 
A more intriguing approach for the generation of structured beams has been developed by Radwell {\it et. al.} in a cold rubidium system \cite{Radwell}.
They have used a single phase structured light beam \cite{Marrucci} and static magnetic field to form a closed-loop Hanle configuration  \cite{Renzoni} in a four level atomic system.
The relative phase difference between the applied fields can drastically modify the Zeeman coherences of a closed-loop transition \cite{Buckle,Kosachiov}.
The phase-dependent Zeeman coherence is a basic ingredient in the control of optical dispersion, absorption, and nonlinearity \cite{Agarwal,Li_2009,Kosachiov_2000,Rajitha,Davuluri,Li_2010,Eilam_2009,Mair_2002}.
Manipulation of these coherences along the azimuthal plane is the main key behind spatially dependent electromagnetic transparency.
Hence, an opaque medium becomes transparent at certain angular positions due to the presence of a phase structured beam. 
Thus controlling transparency in the transverse direction creates a new avenue for the generation of the structured beam.
Radwell {\it et. al.} \cite{Radwell} used a basic theoretical model based on Fermi-golden rule and provided an approximate expression for the periodic variation of the absorption profile to demonstrate how the structured beam can be produced. 
However to achieve good agreement with experiments, various transverse and longitudinal relaxation effects must be incorporated in the propagation dynamics of the light beam with an azimuthally varying polarization and phase structure.

In this paper, we provide a detailed theoretical explanation for the recent experiments on the generation of structured beams \cite{Radwell}, based on full density matrix equations.
To facilitate these structured beam generation, we use a homogeneously broadened four level atomic system driven by two orthogonal polarization components of probe beam as shown in Fig.~\ref{fig:Fig1}.
In order to create phase-dependent atomic coherences, we use a weak magnetic field to couple the ground states.   
We start by deriving an analytical expression for the probe susceptibility in the weak field regime. 
However, numerical solutions of density matrix equations at steady state limits is inevitable to obtain the response of the medium at strong probe field intensities.
To illustrate the effect of phase dependent behaviour of the susceptibility on the probe beam propagation, we numerically study paraxial propagation equations.
We find that the phase dependent absorption creates petal like structures on the probe beam. 
The contrast of the generated structured beam can be enhanced by increasing the coupling strength of the lower level magnetic field. 
Furthermore we study the refractive index profile of two orthogonal polarisation components in the presence of strong probe field. 
A high contrast waveguide and anti-waveguide like structure is achieved unlike in the case of the weak field regime. 
We exploit these waveguide features to generate a diffraction controlled high contrast petal-like beam structure. 
Finally we show the rotation of the generated petal-like beam structure due to magneto optical rotation. 

The paper is organised as follows.
In the next section, we introduce the theoretical model and derive the effective Hamiltonian for a four-level closed-loop atomic configuration.
In Sec. II.B, we adopt density matrix formalism to study the evolution of the atomic population and coherences.
In Sec. II.C, we analytically derive the linear atomic responses to the orthogonal polarization components of the probe field.
In Sec. II.D, we describe the paraxial beam propagation equation for the spatial evolution of phase structured probe field.
Next we provide the results on azimuthally varying susceptibilities for both weak and strong probe field regimes. 
Finally our numerical results delineate the effect of the linear and nonlinear susceptibilities on the propagation dynamics of the probe beam. 
Sec. IV provides a summary of our work.
\section{Theoretical Formulations}
\subsection{Model}
\begin{figure}[t]
\includegraphics[width=8cm,height=8cm,keepaspectratio]{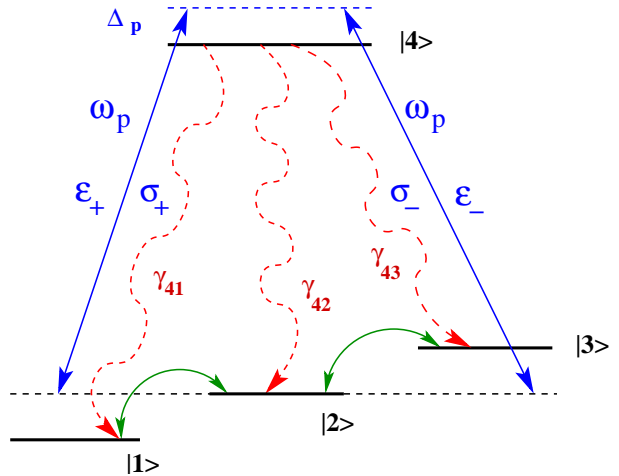}
\caption{\label{fig:Fig1} (Color online) 
Schematic diagram of the four-level closed atomic system. The atomic transition $\ket{4}\leftrightarrow\ket{3}$ and $\ket{4}\leftrightarrow\ket{1}$ are coupled by left($\hat{\sigma}_{-}$) and right($\hat{\sigma}_{+}$) circularly polarized component of the probe field,respectively, whereas the Zeeman sub-levels $\ket{1}$, $\ket{2}$, and $\ket{3}$ are coupled by a transverse magnetic field. $\gamma_{4i}$ corresponds to the radiative decay rates from excited state $\ket{4}$ to ground states $\ket{i}$ where $i\in1,2,3$.}
\end{figure}

The system under consideration is shown in Fig.~\ref{fig:Fig1} where the electric dipole allowed transitions $\ket{1}\leftrightarrow\ket{4}$ and $\ket{3}\leftrightarrow\ket{4}$ are coupled by two orthogonal polarisations $\hat{\sigma}_{+}$ and $\hat{\sigma}_{-}$ of  the probe field, respectively. 
Thus the electric field propagation along the $z$-axis, containing both orthogonal polarisations  with carrier frequency $\omega_p$, can be written as
\begin{equation}
\label{field}
 {\vec{E}}(\vec{r},t)= (\hat{\sigma}_{+}\mathcal{E}_{+}(\vec{r})+\hat{\sigma}_{-}\mathcal{E}_{-}(\vec{r}))~e^{- i\left(\omega_p t-  k_p z\right )} + {c.c.}\,,
\end{equation}
where, $\mathcal{E}_{+}(\vec{r})$ and $\mathcal{E}_{-}(\vec{r})$ are the slowly varying envelopes of right and left circularly polarized probe fields, respectively. The  wave number of probe field is denoted by $k_p$.
An arbitrary magnetic field $\vec{B}= B(\cos\theta~\bf{\hat{z}}+\sin\theta~\bf{\hat{x}})$ is used to connect  
the electric dipole forbidden transitions $\ket{1}\leftrightarrow\ket{2}$ and $\ket{2}\leftrightarrow\ket{3}$ in order to form a closed loop system. 
Such a closed loop system exhibits interesting phase dependent behaviour of absorption and dispersion. 
The longitudinal component of the magnetic field $B\cos\theta$ induces the Zeeman shift between the states $\ket{1}$, and $\ket{3}$   whereas the transverse component $B\sin\theta$ can be used to redistribute the population among the ground states $\ket{1}$, $\ket{2}$, and $\ket{3}$. 
This level scheme has been realised experimentally in cold atomic $^{87}Rb$ vapour where the ground levels $\ket{1}=\ket{5^{2}S_{1/2},F=1,m_{F}=-1}$, $\ket{2}=\ket{5^{2}S_{1/2},F=1,m_{F}=0}$, $\ket{3}=\ket{5^{2}S_{1/2},F=1,m_{F}=1}$, and the excited level $\ket{4}=\ket{5^{2}P_{3/2},F^{'}=0,m^{'}_{F}=0}$.
In the presence of probe and magnetic  fields, the Hamiltonian of the system in the approximations of electric dipole and rotating wave takes the following form
\begin{subequations}
\label{Hschroed}
\begin{align}
H =& H_0 + H_I + H_B\,,\\
H_0 =& \hbar\omega_{42}\ket{4}\bra{4}\,,\\
H_I =& -\hat{D}\cdot\hat{E} \nonumber \\  
    =& -\hbar( \ket{4}\bra{1} g_{1}e^{- i\omega_p t} + \ket{4}\bra{3} g_{2}e^{- i\omega_p t}\,+\,\text{H.c.})\,,\\
H_B =& g_{F}\mu_{B}\hat{F}\cdot\vec{B} \nonumber \\
    =& \hbar\beta_{L}(\ket{3}\bra{3}-\ket{1}\bra{1})+ \hbar\beta_{T}(\ket{1}\bra{2}+\ket{2}\bra{3}\,+\,\text{H.c.}),
 \end{align}
\end{subequations}
where 
\begin{equation}
\label{field}
g_{1}=\frac{\vec{d}_{+}\cdot\vec{\mathcal{E}}_{\rm{+}}}{\hbar}e^{ik_p z}~~\textrm{and}~~g_{2}=\frac{\vec{d}_{-}\cdot\vec{\mathcal{E}}_{\rm{-}}}{\hbar}e^{ik_p z},\nonumber
\end{equation}
are the Rabi frequencies of the probe fields corresponding to the left and right circular polarizations, respectively.
The magnitude of Zeeman shift and the coupling strength between the ground levels are given by $\beta_{L}=g_{F}\mu_{B}B\cos\theta$  and $\beta_{T}=g_{F}\mu_{B}B\sin\theta/\sqrt{2}$, respectively.
We use following unitary transformation 
\begin{equation}
\label{unitary}
W=e^{-\frac{i}{\hbar}Ut}~~\textrm{where}~~U=\hbar\omega_p\ket{4}\bra{4},\nonumber
\end{equation}
to express the Hamiltonian in the time independent form as given below
\begin{eqnarray}
\label{interaction_hamiltonian}
&H_{I}=WHW^{\dagger}=\hbar\Delta_p\ket{4}\bra{4} -\hbar\left(g_{1}\ket{4}\bra{1} +g_{2}\ket{4}\bra{3}\right )\nonumber\\
&+\hbar\beta_{L}(\ket{3}\bra{3}-\ket{1}\bra{1})+ \hbar\beta_{T}(\ket{1}\bra{2}+\ket{2}\bra{3})+\text{H.c.},
\end{eqnarray}
where $\Delta_p=\omega_p-\omega_{42}$ is the probe detuning.
\subsection{Equation of motion}
We now present the full density matrix formalism to study  the experimental work by  Radwell et al.\cite{Radwell}.
The closed loop tripod system possesses various radiative and non-radiative processes. 
To account for these incoherent decay, we use following the Liouville equation
\begin{align}
\label{master}
\dot{\rho}=-\frac{\ii}{\hbar}\left[H_{I},\rho\right]+\mathcal{L}\rho\,.
\end{align}
The second term in Eq.(\ref{master}) represents radiative processes and non-radiative processes that can be determined by
\begin{align}
\mathcal{L}\rho = \mathcal{L}_{r}\rho+\mathcal{L}_{c}\rho
 \label{decay1}
\end{align}
with
\begin{subequations}
\begin{align}
\mathcal{L}_{r}\rho = &-\sum\limits_{i=1}^3 \frac{\gamma_{4i}}{2}\left(\ket{4}\bra{4}\rho-2\ket{i}\bra{i}\rho_{44}+\rho\ket{4}\bra{4}\right)\,,\nonumber\\
\mathcal{L}_{c}\rho = &-\sum\limits_{j=1}^3\sum\limits_{j\ne i=1}^3 \frac{\gamma_{c}}{2}\left(\ket{j}\bra{j}\rho-2\ket{i}\bra{i}\rho_{jj}+\rho\ket{j}\bra{j}\right)\,\nonumber.
 \label{decay2}
\end{align}
\end{subequations}
The first term of Eq.(\ref{decay1}) represents radiative decay from excited state $\ket{4}$ to ground states $\ket{i}$, and are labeled by $\gamma_{4i} (i\in1,2,3)$,  whereas
the second term represents pure dephasing for the coherence $\rho_{ij}$ due to collision at a rate $\gamma_c$.
The dynamics of the atomic population and coherences for the closed loop tripod system can be obtained by substituting the effective Hamiltonian
(\ref{interaction_hamiltonian}) in the Liouville equation (\ref{master}). 
Therefore, the following set of Bloch equations can be conveniently written
\begin{subequations}
\label{Full_density}
\begin{align}
 \dot{\rho}_{11}=&\gamma_{41}\rho_{44}-\ii\beta_{T}\rho_{21}+\ii\beta_{T}\rho_{12}+ \ii g^{*}_{1} \rho_{41} \nonumber \\
& - \ii g_{1} \rho_{14} - \gamma_{c}\rho_{11}+ \gamma_{c}\rho_{22}+ \gamma_{c}\rho_{33} \,,\\
 \dot{\rho}_{12}=&\ii\beta_{L}\rho_{12}-\ii\beta_{T}(\rho_{22}-\rho_{11})+\ii\beta_{T}\rho_{13}  \nonumber \\
& + \ii g^{*}_{1} \rho_{42}- \gamma_{c}\rho_{12}\,, \\
 \dot{\rho}_{13}=&2\ii\beta_{L}\rho_{13}-\ii\beta_{T}(\rho_{23}-\rho_{12})+ \ii g^{*}_{1} \rho_{43}  \nonumber \\
& - \ii g_{2} \rho_{14} - \gamma_{c}\rho_{13}\,,\\
 \dot{\rho}_{14}=&-\ii(\Delta_{p}-\beta_{L})\rho_{14}-\ii\beta_{T}\rho_{24}+\ii g^{*}_{1}(\rho_{44}-\rho_{11})  \nonumber \\
& - \ii g^{*}_{2} \rho_{13} - (\Gamma_{41}+\gamma_{p}) \rho_{14} \,,\\
 \dot{\rho}_{22}=&\gamma_{42}\rho_{44} -\ii\beta_{T}(\rho_{12}-\rho_{21})-\ii\beta_{T}(\rho_{32}-\rho_{23}) \nonumber \\
& + \gamma_{c}\rho_{11}- \gamma_{c}\rho_{22}+ \gamma_{c}\rho_{33}\,,\\
 \dot{\rho}_{23}=&\ii \beta_{L}\rho_{23}-\ii \beta_{T}\rho_{13}-\ii \beta_{T}(\rho_{33}-\rho_{22})  \nonumber \\
&- \ii g_{2} \rho_{24}- \gamma_{c}\rho_{23}\,,\\
 \dot{\rho}_{24}=&-\ii \Delta_{p}\rho_{24}-\ii \beta_{T}(\rho_{14}+\rho_{34})-\ii g^{*}_{1}\rho_{21} \nonumber \\
& - (\Gamma_{42}+\gamma_{p}) \rho_{24} -\ii g^{*}_{2}\rho_{23}\,,\\
\dot{\rho}_{33}=&\gamma_{43}\rho_{44} -\ii\beta_{T}(\rho_{23}-\rho_{32})+ \ii g^{*}_{2} \rho_{43} \nonumber \\
& + \ii g_{2} \rho_{43} + \gamma_{c}\rho_{11}+ \gamma_{c}\rho_{22}- \gamma_{c}\rho_{33}\,,\\
 \dot{\rho}_{34}=&-\ii(\Delta_{p}+\beta_{L})\rho_{34}-\ii\beta_{T}\rho_{24}+\ii g^{*}_{2}(\rho_{44}-\rho_{33}) \nonumber \\
& - \ii g^{*}_{1} \rho_{31} - (\Gamma_{43} + \gamma_{p}) \rho_{34} \,.
 \end{align}
\end{subequations}
The remaining density matrix equations come from the population conservation law $\sum\limits_{i=1}^{4}\rho_{ii}=1$ and the complex conjugate expressions $\dot{\rho}_{ji}=\dot{\rho}_{ij}^*$.
\subsection{Probe susceptibility of a homogeneous medium}
In this section, we calculate the linear response of the probe field in a homogeneous medium.
The probe field is to be weak enough to be treated as a perturbation to a system of linear order under steady-state condition.
This assumption leads us  to get a good agreement of the recent experiment results \cite{Radwell}.
The perturbative expansion of the density matrix upto first order of probe field $g_i, (i\in 1,2)$ can be expressed as
  \begin{equation}
  \label{perturb}
  \rho_{_{ij}}=\rho_{_{ij}}^{(0)}+g_{_{1}}\rho_{_{ij}}^{(+)}+g_{_{2}}\rho_{_{ij}}^{(+)},
  \end{equation}
where, $\rho_{ij}^{(0)}$ is the solution in the absence of the probe field.
The second and third terms in Eq.(\ref{perturb}) denote first-order solutions of the density matrix elements for both orthogonal polarizations at positive probe field frequency $\omega_p$.
We now substitute Eq.~(\ref{perturb}) in Eqs (\ref{Full_density}) and equate the coefficients of $g_{1}$ and $g_{2}$. 
As a result, we obtain two sets of 12 coupled linear equations. 
Next, we solve these algebraic equations to derive the atomic coherences $\rho_{_{41}}^{(+)}$ and $\rho_{_{43}}^{(+)}$.
The  off-diagonal density matrix elements $\rho_{_{41}}^{(+)}$ and $\rho_{_{43}}^{(+)}$ determine the linear susceptibility $\chi_{_{41}}$ and $\chi_{_{43}}$ of the medium at frequency $\omega_{p}$ respectively. 
Hence the medium polarization induced by the probe field can be expressed as 
\begin{subequations}
\begin{align}
\label{chi_41-43}
{\chi}_{_{41}}(\DP)&=\frac{\mathcal{N}|d_{+}|^2}{{\hbar}}{\rho}_{41}^{(+)}\,, \\
{\chi}_{_{43}}(\DP)&=\frac{\mathcal{N}|d_{-}|^2}{{\hbar}}{\rho}_{43}^{(+)},
\end{align}
\end{subequations}
with 
\begin{align}
{\rho}_{41}^{(+)}=& \frac{N_{1}}{D}g_{1}+\beta^{2}_{T}\frac{N_{2}}{D}g_{2}\,, \\
{\rho}_{43}^{(+)}=& \beta^{2}_{T}\frac{N_{2}}{D}g_{1}+\frac{N_{3}}{D}g_{2}\,,
\end{align}
where
\begin{align}
N_{1}=& \bl^{2}(\DP + i\Gamma_{41})(\DP + i\Gamma_{41}+ \bl)\rho^{0}_{11} \nonumber \\
&+\bs^{4}(2\rho^{0}_{11} - \rho^{0}_{22})- \bs^{2}(\rho^{0}_{11}(\DP + i\Gamma_{41})^{2} \nonumber \\
& +\bl\rho^{0}_{22}(\DP + i\Gamma_{41}+ \bl))\,, \\
N_{2}=& (\DP + i\Gamma_{41})^{2}(\rho^{0}_{11} - \rho^{0}_{22})+\bl^{2}\rho^{0}_{22} \nonumber \\
& -\bs^{2}(2\rho^{0}_{11} - \rho^{0}_{22})\,, \\
N_{3}=& \bl^{2}(\DP + i\Gamma_{41})(\DP + i\Gamma_{41}- \bl)\rho^{0}_{11} \nonumber \\
& +\bs^{4}(2\rho^{0}_{11} - \rho^{0}_{22})-\bs^{2}(\rho^{0}_{11}(\DP + i\Gamma_{41})^{2} \nonumber \\
& +\bl\rho^{0}_{22}(\DP + i\Gamma_{41}- \bl))\,, \\
D=& (\bl^{2}-\bs^{2})(\bl^{2}+2\bs^{2}-(\DP + i\Gamma_{41})^{2}) \nonumber \\
&(\DP + i\Gamma_{41}) \nonumber\\
\label{pop1}
\rho^{0}_{11}=& (\frac{\bs}{\bl})^{2} \,,\\
\label{pop2}
\rho^{0}_{22}=& 1-2(\frac{\bs}{\bl})^{2} \,.
\end{align}
Here $\mathcal{N}$ is the atomic density of the medium.
The influence of transverse magnetic field on the steady state population in absence of the probe field is clearly seen from Eqs.(\ref{pop1}) and (\ref{pop2}).
The phase dependent response of the medium can be explored by considering the spatial inhomogeneity of the probe field.
Thus the spatial structure of the probe field for two orthogonal polarisations can be expressed as
\begin{align}
g_{1}(r,\phi)=& g(r)e^{il\phi}\,, \\
g_{2}(r,\phi)=& g(r)e^{-il\phi}\,,
\end{align}
where $l$, $\phi$ and $g(r)$ represents OAM, phase and transverse variation of the probe beam, respectively.
The phase dependent susceptibilities of the closed loop tripod system is given as
\begin{align}
\chi_{_{41}}=& \frac{\mathcal{N}|d_{+}|^2}{{\hbar}}\left(\frac{N_{1}}{D}+\beta^{2}_{T}e^{-2il\phi}\frac{N_{2}}{D}\right)\label{chi_41}\,, \\ 
\chi_{_{43}}=& \frac{\mathcal{N}|d_{-}|^2}{{\hbar}}\left(\frac{N_{3}}{D}+\beta^{2}_{T}e^{2il\phi}\frac{N_{2}}{D}\right)\label{chi_43}\,. 
\end{align}
The above analytical expressions of the susceptibilities for the transitions $\ket{1}\leftrightarrow\ket{4}$ and $\ket{3}\leftrightarrow\ket{4}$ display an insight on the physics behind the formation of structured beam profile.
The transverse magnetic field $\beta_T$, OAM $l$ and transverse phase $\phi$ plays a crucial role in the manipulation of  the optical properties of closed loop tripod systems.
We adopt Gauss-Jordan elimination method to solve linear algebraic Eq.~(\ref{Full_density}) numerically at steady-state condition of the density matrix for a probe field at higher intensities limits.
\subsection{Beam propagation equation with paraxial approximation}
In order to investigate the effect of azimuthally varying susceptibilities on both left and right polarized components of the probe beam, we use Maxwell's wave equations under slowly varying envelope and paraxial wave approximations.
The dynamics of the orthogonal polarization components with Rabi frequencies $g_{1}$ and $g_{2}$ propagating along the $z$-direction
can be expressed in the following form:  
\begin{subequations}
\label{c4}
\begin{align}
 \frac{\partial g_{1}}{\partial z}
   &= \frac{\ii }{2{k_p}} \left( \frac{\partial^2 }{\partial x^2}
      + \frac{\partial^2 }{\partial y^2} \right) g_{1} + 2i{\pi}k_p{\chi}_{41}\,{g}_{1} \,,\label{probe} \\
 \frac{\partial g_{2}}{\partial z}
   &= \frac{\ii }{2{k_p}} \left( \frac{\partial^2 }{\partial x^2}
      + \frac{\partial^2 }{\partial y^2} \right) g_{2} + 2i{\pi}k_p{\chi}_{43}\,{g}_{2} \,.\label{control} \,
\end{align}
\end{subequations}
The terms within the parentheses on the right hand side of Eq.~(\ref{probe}) and  Eq.~(\ref{control}) are account for transverse variation of the probe beam. 
These terms responsible for the diffraction either in the medium or in free space. 
The second terms on the right-hand side of Eq.~(\ref{probe}) and  Eq.~(\ref{control}) leads to the dispersion and absorption of the probe beam.
\begin{figure}[b]
\includegraphics[width=8cm,height=8cm,keepaspectratio]{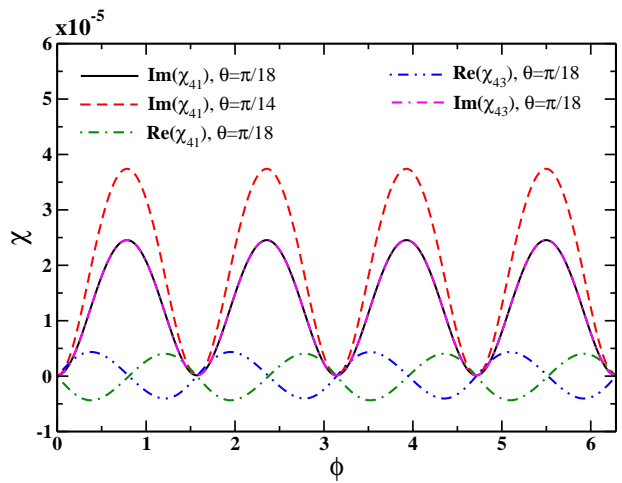}
\caption{\label{fig:Fig2} (Color online) 
Real and imaginary part of susceptibilities $\chi_{41}$ and $\chi_{43}$ as a function of phase for different $\theta$ are plotted. The parameters are chosen as ${\mathcal{N}}=10^{11} atoms/cm^{3}$, $\Gamma_{41}=0.5{\gamma}$, $\DP =0$, $\beta_{0}=0.01{\gamma}$, $\gamma_{c}=10^{-7}{\gamma}$, $g_{0}=0.01\gamma$ and $l=2$.}
\end{figure}
\section{Results and Discussions }
\subsection{Azimuthally varying susceptibility}
\begin{figure}[t!]
\includegraphics[width=8.5cm, height=7.0cm,keepaspectratio]{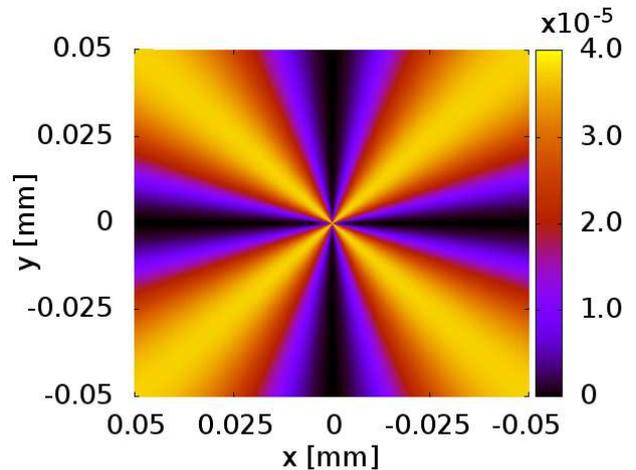}
\caption{\label{fig:Fig3} (Color online) 
Absorption pattern of the $\hat{\sigma}_{+}$ polarization component is plotted against the two orthogonal axes $x$ and $y$. Other parameters are same as in Fig.~\ref{fig:Fig2}.}
\end{figure}

We first study the effect of azimuthal phase on the absorption of two orthogonal polarisation components $\hat{\sigma}_{\pm}$ of the probe field at weak intensity regime.
The phase dependent susceptibilities $\chi_{41}$ and $\chi_{43}$ can be explored by considering the amplitude of both the polarization components to be continuous wave with $g_{i}(r)=g_0=0.01\gamma, (i\in1,2)$. 
We assume the collisional decay term $\gamma_{c}$ is very negligible to be consistent with the experimental results for the cold atomic system \cite{Radwell}.
The absorption of right- and left-handed circular polarizations $\chi_{41}$ and $\chi_{43}$ are plotted against the azimuthal phase as shown in Fig.~\ref{fig:Fig2}.
Results are presented in Fig.~\ref{fig:Fig2} for two different intensities of magnetic field.
From Fig.~\ref{fig:Fig2}, we find that the absorption at the $\hat{\sigma}_{+}$ transition oscillates periodically.
The periodic variation of this absorption can be well explained by considering the perturbative expression for the susceptibility 
as mentioned in Eq.~(\ref{chi_41}).
The term associated with the second fraction in the round bracket of Eq.~(\ref{chi_41}) leads to phase dependent response of the medium.
The $2l$ factor in the exponential term decides the number of transparency windows that can be formed within a period.
It is clear from Fig.~\ref{fig:Fig2} that OAM $l=2$ creates $4$ transparency windows.
The narrowing of the transparency window is a key mechanism to generate the high contrast periodic absorption structures.
It is evident from Fig.~\ref{fig:Fig2} that the sharp variation of transparency windows can be achieved by increasing $\theta$ value.
The increment of $\theta$ returns the strength of TMF $\bs$ at higher values that leads to an increase in the population of the ground states $\ket{1}$, and $\ket{3}$ as shown by the Eqs.(\ref{pop1}) and (\ref{pop2}).  
As a result, each polarisation component suffer more absorption due to the narrowing of the transparency window.
Hence the strength of TMF  and the azimuthal phase plays an important role in creating a  high contrast periodic absorption structure for the probe field.
Note that the absorption of the left-handed polarisation $\hat{\sigma}_{-}$ is identical to the absorption of the right-handed polarisation $\hat{\sigma}_{+}$  at $\Delta_p=0$ as shown in Fig.~\ref{fig:Fig2}.
The absorption of  both circularly polarised components constitute the structure of the probe absorption.
The phase dependent absorption structure is a result of the coupling among the degenerate ground states by a weak magnetic field.
The degeneracy between the ground states $\ket{i}, (i\in1,2,3)$ can be lifted in the presence of a longitudinal magnetic field $\beta_L$.
The probe resonance condition $\Delta_p=0$ facilitates the red and blue shifted detuning by an amount of $\beta_L$ for each  circularly polarised component.
Thus the refractive index profile for $\hat{\sigma}_{+}$ component varies oppositely as the refractive index profile for $\hat{\sigma}_{-}$ component with a very small magnitude as shown in Fig.~\ref{fig:Fig2}.
This reverse nature of refractive index for both polarization components failed to  resemble the wave-guided structure inside the medium.
Hence the created structure of the probe beam suffers distortion due to diffraction.
In Fig.~\ref{fig:Fig3}, we show the surface plot of $\chi_{41}$ as a function of transverse directions $x$ and $y$.
Two orthogonal axes $x$ and $y$ can be used to define azimuthal phase $\phi=\textrm{tan}^{-1}(y/x)$. 
It can be seen from Fig.~\ref{fig:Fig3} that the medium becomes transparent at some specific angular positions for the right-handed polarisation.
These angular positions can be defined by $n\pi/l$ where $n$ can change from $0~\textrm{to}~l$.
As a result of angular dependency, the absorption profile for the right-handed polarisation shows fourfold symmetry with OAM, $l=2$.  
A similar periodic absorption pattern is exhibited by the left-handed circularly polarised component as mentioned in Eq.~(\ref{chi_43}).
Note that in absence of phase modulation, both polarisation components suffer from high attenuation.
Thus at weak field limits, in a closed loop tripod system, the phase information of each polarisation component gets converted into the intensity information that renders transparent an otherwise opaque medium.
\begin{figure}[t!]
\includegraphics[width=8cm,height=8cm,keepaspectratio]{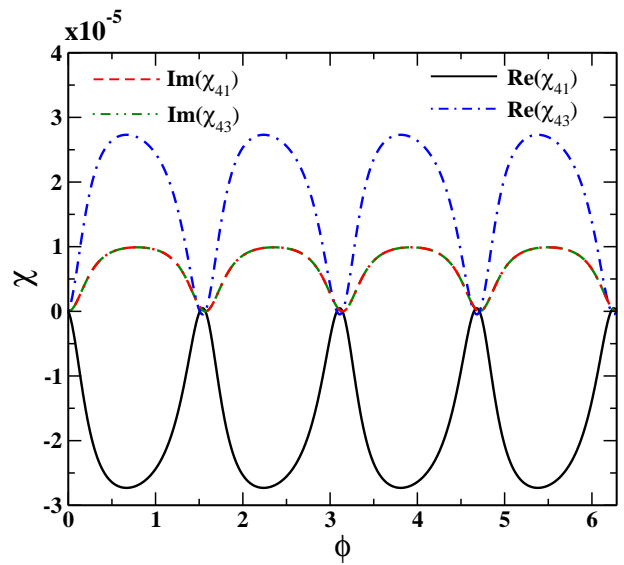}
\caption{\label{fig:Fig4} (Color online) 
The variations of real and imaginary part of susceptibilities $\chi_{41}$ and $\chi_{43}$ as a function of phase for relatively strong probe regime are plotted. The parameters are chosen as ${\mathcal{N}}=10^{11} atoms/cm^{3}$, $\Gamma_{41}=0.5{\gamma}$, $\DP =0$, $\beta_{0}=0.01{\gamma}$, $\theta=\pi/14$, $\gamma_{c}=10^{-7}{\gamma}$ and $l=2$.}
\end{figure}
\begin{figure}[t!]
\subfigure[]
 {
 \centering
   \includegraphics[width=8.5cm, height=7.0cm]{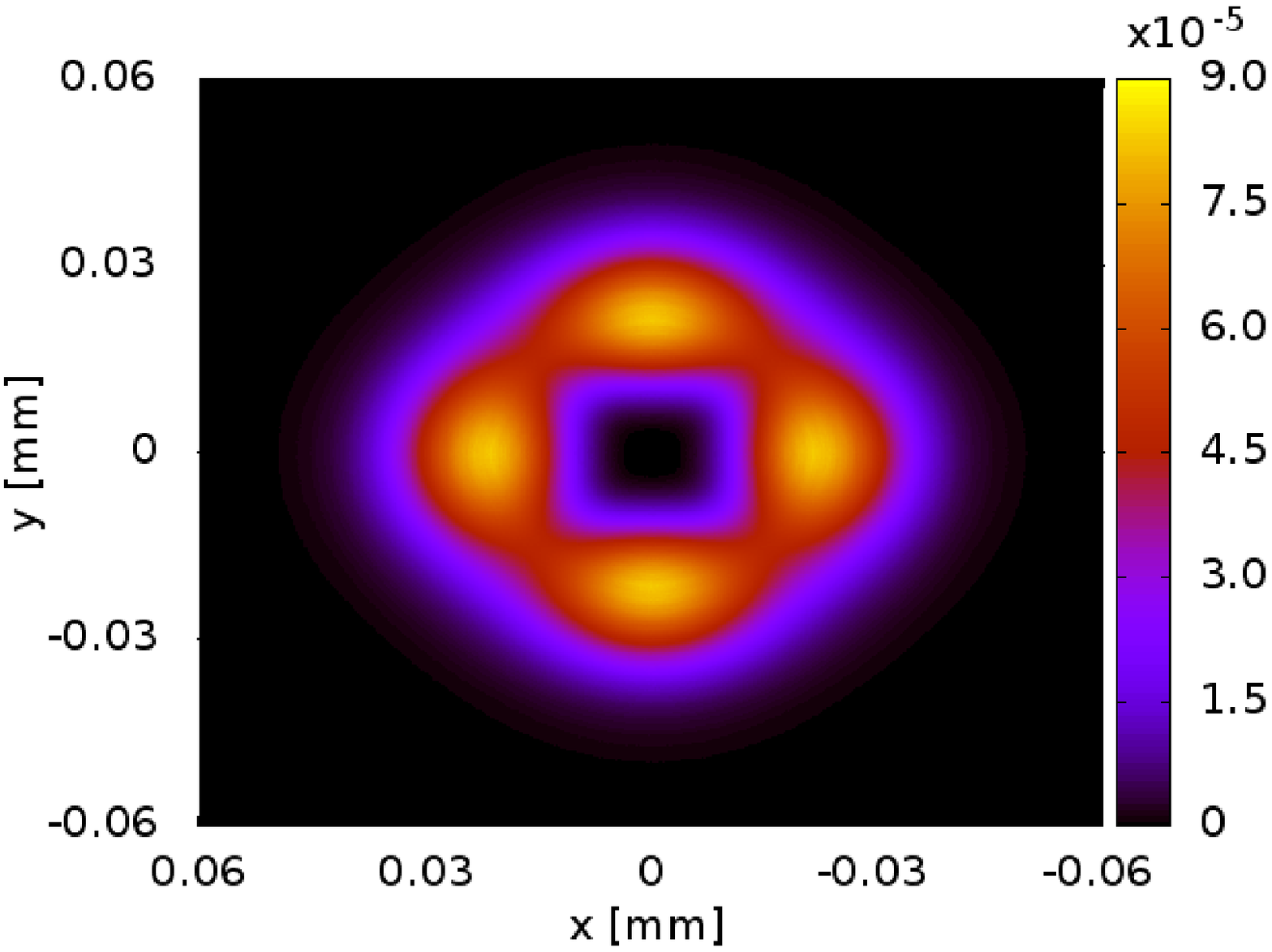}
   \label{fig:Fig7a}
 }
\subfigure[]
 {
 \centering
 \includegraphics[width=8.5cm, height=7.2cm]{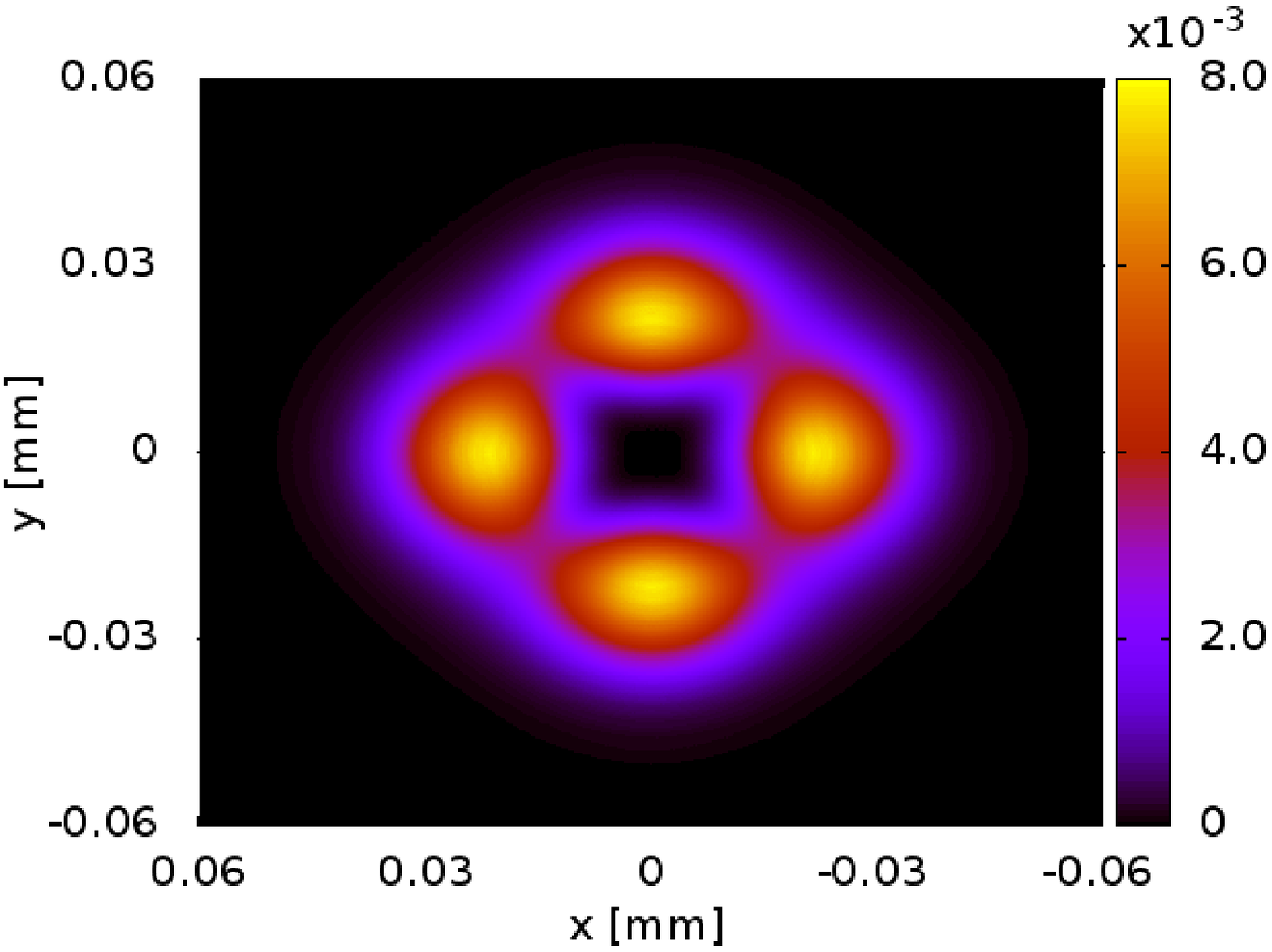}
   \label{fig:Fig7b}
 }
\caption{\label{fig:Fig5} (Color online) Panel (a) and (b) depicts transmitted probe beam intensity in the transverse $(x-y)$ plane for $\theta=\pi/18$ and  $\pi/14$, respectively. 
The intensity profile of the probe beam is shown in the panel (a) and (b) after traverse a distance of medium length $ 0.6~mm$.
The mode, OAM and waist of the Laguerre-Gaussian beam are $m=0$, $l=2$ and $w_{p}=20 \mu$m, respectively at $z=0$.
Other parameters are same as in Fig.~\ref{fig:Fig2}.}
\end{figure}

We now discuss the response of the medium beyond the weak field limits as shown in Fig.(\ref{fig:Fig4}).
For a relatively strong probe field limit $g_{0}=0.1\gamma$,  the numerical solutions of  linear algebraic equations (\ref{Full_density}) are inevitable to analyze the phase dependent  susceptibility of the medium at steady state condition.
The oscillating amplitude of polarization components reduces with increase in $g_{0}$.
As a consequence the population in the ground states $\ket{1}$, and $\ket{3}$ gets depleted. 
The depletion of population in these ground states is the cause of  width-broadening of the transparency window.
Surprisingly the refractive index profile of two orthogonal polarisation components modify drastically as compared to the case in a weak field regime.
It is evident from  Fig.~\ref{fig:Fig4} that the gradient of refractive index is dependent on the detuning sign of polarization components.
The slope of the refractive index attains its maximum around the transparency window and decreases gradually towards the wings for a red detuned right circular polarisation component. 
 A convex lens like refractive index is formed for the red shifted polarisation component whereas concave lens like refractive index
 is experienced by the blue shifted polarisation component.
Thus by selecting detuning of two orthogonal polarisations $\hat{\sigma}_{\pm}$, leads to the formation of  a waveguide \cite{Truscott} and an anti-waveguide \cite{Bortman} in the closed loop tripod system. 
Hence these waveguide/antiwaveguide structures can lead to focusing/defocusing of polarisation components. 
These features are missing for the weak intensity limits as the susceptibilities are independent of polarisation amplitude as shown in Eq.~(\ref{chi_41}).
Also the amplitude of the refractive index is stronger here than in the weak field limits.
Hence a suitable choice of detuning of each polarization component at the strong field regime can lead to diffraction controlled petal like structured beam generation.
\subsection{Beam propagation dynamics}
\begin{figure}[t!]
\includegraphics[width=8.2cm, height=6.9cm]{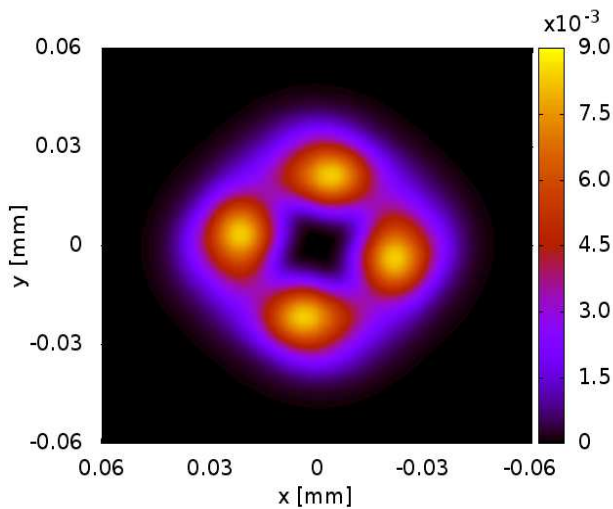}
\caption{\label{fig:Fig6} (Color online) 
Intensity variation of the probe in transverse $(x-y)$ plane, after propagating through a medium of $ 0.6~$mm long. 
The initial amplitude, mode, OAM and waist of the Laguerre-Gaussian beam are $g_{0}=0.1 \gamma $, $m=0$, $l=2$ and $w_{p}=20 \mu$m, respectively.
Other parameters are same as in Fig.~\ref{fig:Fig4}.}
\end{figure}

Next we illustrate how spatially dependent susceptibility enables us to generate the structured probe beam.
For this purpose, the transverse spatial profile of both the polarisation components is to be in the Laguerre-Gaussian mode that can be written as
\begin{align}
g_{j}(r,z)=& g_{0}\times\frac{w_{p}}{w(z)}\times\left(\frac{r\sqrt{2}}{w(z)}\right)^{\left|l\right|}L^{l}_{m}\left(\frac{2r^{2}}{w^{2}(z)}\right)\times e^{\pm i l\phi}\nonumber \\
     & e^{-\left(\frac{r^{2}}{w^{2}(z)}\right)}e^{\left(\frac{ikr^{2}}{2R(z)}\right)}e^{-i(2m+l+1)\tan^{-1}\left(\frac{z}{z_{0}}\right)}\, \\
r=& \sqrt{x^{2}+y^{2}}\,\nonumber \\
\phi=& \tan^{-1}\left(\frac{y}{x}\right).\nonumber
\end{align}
The indices $m$ determine the shape of the probe field profile along the transverse directions.
The radius of curvature and the Rayleigh length are defined as $R(z)=z+(z^{2}_{0}/z)$, and $z_{0}=\pi w^{2}_{p}/\lambda$, respectively. 
The beam width is varied with propagation distance $z$ as $w(z)=w_{p}\sqrt{1+(z/z_{0})^{2}}$, where, $w_{p}$ is the beam waist at $z=0$ \cite{Milonni}.
We adopt higher ordered split-step operator method to numerically study the beam propagation Eq.~(\ref{c4}). 
Fig.~\ref{fig:Fig5} shows the output intensity pattern of the probe beam at a propagation distance of $0.6~$mm.
Since the left and right-handed polarization components are orthogonal to each other, therefore the output intensity can be expressed as  $I_{out}= \left|g_{1}\right|^{2} + \left|g_{2}\right|^{2}$.
It is evident from Fig.~\ref{fig:Fig5} that the fourfold symmetry which exists in the absorption profile of both the polarization components are mapped onto their transverse spatial profile.
Also this spatial profile assures that each polarization component carries OAM with units of $\pm2\hbar$.
Thus the value of OAM dictates the formation of structured probe beam with desired shape. 
Hence manipulation of the absorption profile along the transverse direction forms the key idea behind the structured probe beam generation. 
We also find that the transmission for the probe beam at a propagation distance of $0.6~$mm is 71$\%$ for $\theta=\pi/18$ and 62$\%$ for $\theta=\pi/14$ even at weak field limits.
We further study how the magnetic field strength allows us to enhance the contrast of the structured beam. 
A higher strength $\bs$ creates a high contrast periodic absorption profile as compared with a weak $\bs$ as shown in Fig.~\ref{fig:Fig2}.
This sharp variation of absorption profile emulates a high contrast spatial probe beam profile as depicted in Fig.~\ref{fig:Fig6}(b).
On the other hand, Fig.~\ref{fig:Fig5}(a) shows a low contrast spatial probe beam for the weak field regime at a propagation distance of $z=0.6$ mm.
Thus the contrast enhancement of  the structured beam is possible by using a suitable magnetic field strength. 

We next study the effect of nonlinear susceptibility on the spatial evolution of the probe beam envelope.   
As seen from Fig.~\ref{fig:Fig6}, the output pattern of the probe intensity bears the same four-fold symmetric patterns as in the weak field case. 
However, the contrast and rotation of the patterns are changed significantly as compared to the later. 
Fig.~\ref{fig:Fig4} exhibits the waveguide and anti-waveguide refractive index profile for the constituents of the probe beam at $\Delta_p=0$, which enhances the contrast of the output pattern.
The waveguide structure confine the $\sigma_+$ polarization component whereas the $\sigma_-$ polarization component gets defocused due to the anti-waveguide structure. 
Moreover, the spreading of both the polarization components  in the azimuthal plane is limited by the width of the spatial transparency window.
Hence at the strong regime both absorption and dispersion profiles  play important roles in improving the output beam pattern whereas the absorption profile is solely responsible in the weak field case.
The transmission of the structured beam at a propagation distance of $0.6~mm$ is found to be 66$\%$. 
The increase in beam transmission is due to waveguide induced focusing of the probe beam in the azimuthal plane.  
We also notice from Fig.~\ref{fig:Fig6} that the generated structure beam is rotated by an angle of $10^{\circ}$. 
This rotation is attributed to strong field induced NMOR. 
The rotation of the structured beam can be enhanced by increasing the intensity of the probe and the magnetic field strengths \cite{Budker}.
Our approach opens up new possibilities for generating high contrast structured beam in other closed loop systems that display narrow EIT resonances.
The step variation of refractive index around the narrow transparency window is the main reason behind the formation of high contrast beams.
Thus an atomic medium with buffer gas \cite{Mikhailov} and inhomogeneously broadened atomic system \cite{Dey} may be suitable candidates for creating a diffraction controlled high contrast structured beam.

\section{\label{Conclusion}Conclusion}
In conclusion, we have studied the generation of structured beam in a cold $^{87}$Rb atomic vapor.
For this purpose, a phase dependent transparency is prepared in a closed loop tripod system tailored with a phase structured probe beam and a TMF.
In the presence of the TMF,  the absorption of  both weak and strong fields oscillates periodically in the azimuthal plane. 
However, a waveguide and anti-waveguide like refractive index features are formed for the later case. 
Such a periodic variation of medium absorption and refractive index are responsible for the creation of a high contrast structured beam. 
By numerically solving the propagation equations, we  confirm the formation of petal like beam structure for both weak and strong field cases. 
Interestingly, the features of the generated structured beam is enhanced with high transmittivity at strong field limits. 
Further, increasing the amplitudes of the  phase structured beam and TMF can lead rotation of the petal beam structure  due to NMOR.


\begin{thebibliography}{13}

\bibitem{Woerdemann}M. Woerdemann, C. Alpmann, M. Esseling and C. Denz, Laser Photonics. Rev. {\bf 7}, 839 (2013).

\bibitem{Vaziri}A. Vaziri, J. W. Pan, T. Jennewein, G. Weihs, and A. Zeilinger, \prl {\bf 91}, 227902 (2003).

\bibitem{Gahegan}K. T. Gahegan, and, G. A. Swartzlander, \ol {\bf 21}, 827 (1996).

\bibitem{Padgett}M. Padgett and R. Bowman, Nat. Photon. {\bf 5}, 343 (2011).

\bibitem{Macdonald}M. P. Macdonald, L. Paterson, K. V. Sepulveda, J. Arlt, W. Sibbett, and K. Dholakia, Science, {\bf 296}, 1101 (2002).

\bibitem{Paterson}L. Paterson, M. P. Dacdonald, J. Arlt, W. Sibbett, P. E. Bryant, and K. Dholakia, Science, {\bf 292}, 912 (2001).

\bibitem{Stevenson}D. J. Stevenson, F. G. Moore, and K. Dholakia, Journal of Biomedical optics, {\bf 15}, 041503 (2010).

\bibitem{Zupancic}P. Zupancic, P. M. Preiss, A. Lukin, M. E. Tai, M. Rispoli, R. Islam , and M. Greiner, Opt. Express {\bf 24}, 013881 (2016).

\bibitem{Schwarz}U. T. Schwarz, M. A. Bardres, and J. C. G. Vege, \ol  {\bf 29}, 1870 (2004).

\bibitem{Naidoo}D. Naidoo, K. A. Ameur, M. Brand, and A. Forbes, Appl. Phys. B. {\bf 106}, 683 (2012).

\bibitem{Sakamoto}M. Sakamoto, K. Oka, R. Morita, and N. Murakami, \ol {\bf 38}, 3661 (2013).

\bibitem{Burger}L. Burger, and A. forbes, Opt. Express {\bf 16}, 12707 (2008).

\bibitem{Litvin}I. A. Litvin, L. Burger, and A. forbes, Opt. Express {\bf 15}, 14065 (2007).

\bibitem{Jesacher}A. Jesacher, S. Furhapter, S. Bernet, and M. R. Marte, Opt. Express {\bf 12}, 4129 (2004).

\bibitem{Schmitz}C. H. J. Schmitz, K. Uhrig, J. P. Spatz, and J. E. Curtis, Opt. Express {\bf 14}, 6604 (2006).

\bibitem{Arnold}S. F. Arnold, J. Leach, M. J. Padgett, V. E. Lembessis, D. Ellinas, A. J. Wright, J. M. Girkin, P. Ohberg, and A. S. Arnold, Opt. Express {\bf 15}, 8619 (2007).

\bibitem{Radwell}N. Radwell, T. W. Clark, B. Piccirillo, S. M. Barnett, and S. Franke-Arnold, \prl {\bf 114}, 123603 (2015).

\bibitem{Marrucci}L. Marrucci, C. Manzo, and D. Paparo, \prl {\bf 96}, 163905 (2006).

\bibitem{Renzoni}F. Renzoni, W. Maichen, L. Windholz, and E. Arimondo, \pra {\bf 55}, 3710 (1997).

\bibitem{Buckle} S. J. Buckle, S. M. Barnett, P. L. Knight, M. A. Lauder, and D. T. Pegg, Opt. Acta {\bf 33}, 1129 (1986).

\bibitem{Kosachiov}D. V. Kosachiov, B. G. Matisov, and Y. V. Rozhdestvensky, J. Phys. B {\bf 25}, 2473 (1992).

\bibitem{Agarwal}G. S. Agarwal, T. N. Dey, and S. Menon, \pra {\bf 64}, 053809 (2001).

\bibitem{Li_2009}H. Li, V. A. Sautenkov, Y. V. Rostovtsev, G. R. Welch, P. R. Hemmer, and M. O. Scully, \pra {\bf 80}, 023820 (2009).

\bibitem{Kosachiov_2000}D. V. Kosachiov and E. A. Korsunsky, Eur. Phys. J. D {\bf 11}, 457 (2000).

\bibitem{Rajitha}K. V. Rajitha, T. N. Dey, S. Das, and P. K. Jha, \ol {\bf 40}, 2229 (2015).

\bibitem{Davuluri}S. Davuluri and Y. Rostovtsev, \pra {\bf 88}, 053847 (2013).

\bibitem{Li_2010}L. Li and G. X. Huang, Eur. Phys. J. D {\bf 58}, 339 (2010).

\bibitem{Eilam_2009}A. Eilam, A. D. Wilson-Gordon, and H. Friedmann, \ol {\bf 34}, 1834 (2009).

\bibitem{Mair_2002}A. Mair, J. Hager, D. F. Phillips, R. L. Walsworth, and M. D. Lukin, \pra {\bf 65}, 031802 (2002).

\bibitem{Milonni}P. W. Milonni, J. H. Eberly, Laser Physics, Wiley, 2010.

\bibitem{Truscott}A. G. Truscott, M. E. J. Friese, N. R. Heckenberg, and H. Rubinsztein-Dunlop, \prl {\bf 82}, 1438 (1999).

\bibitem{Bortman}D. Bortman-Arbiv, A. D. Wilson-Gordon, and H. Friedmann, \pra {\bf 63}, 031801(R) (2001).

\bibitem{Budker}D. Budker, D. F. Kimball, V. V. Yashchuk, and M. Zolotorev, \pra {\bf 65}, 055403 (2002).

\bibitem{Mikhailov}E. E. Mikhailov, I. Novikova, Y. V. Rostovtsev, and G. R. Welch, \pra {\bf 70}, 033806 (2004).

\bibitem{Dey}T. N. Dey, and J. Evers, \pra {\bf 84}, 043842 (2011).

\end{thebibliography}
\end{document}